\documentstyle[prc,aps,preprint]{revtex}

\begin{document}
\draft
\tighten

\title{Feynman's ratchet and pawl: an exactly solvable model}
\author{C. Jarzynski$^a$, O. Mazonka$^b$}
\address{~ \\ (a) Theoretical Division, Los Alamos National
                Laboratory, USA \\
         {\tt chrisj@lanl.gov}}
\address{(b) Institute for Nuclear Studies,
                \' Swierk, Poland \\
         {\tt maz@iriss.ipj.gov.pl}}

\maketitle

\begin{abstract}
We introduce a simple, discrete model of Feynman's
ratchet and pawl, operating between two heat reservoirs.
We solve exactly for the steady-state directed motion and 
heat flows produced, first in the absence and then in 
the presence of an external load.
We show that the model can act both as a heat engine
and as a refrigerator.
We finally investigate the behavior of the system near
equilibrium, and use our model to confirm general predictions
based on linear response theory.
\end{abstract}

\pacs{\\
      Keywords: {\bf thermal ratchets} \\
      PACS: 05.40.-a \\ \\
      LAUR-99-568}

\section*{Introduction}
\label{sec:intro}

Feynman's {\it ratchet and pawl} system\cite{feynman}
is a well known (but not the earliest!\cite{smol})
example of a proposed ``mechanical Maxwell's demon'', a device
whose purpose is to convert into useful work the thermal
motions present in a heat reservoir.
The idea is beautifully simple:
set up a ratchet and pawl so that a wheel is allowed to
turn only in one direction, then attach that wheel
to a windmill whose vanes are surrounded by a
gas at a finite temperature; see Fig.46-1 of
Ref.\cite{feynman}.
Every so often, an accumulation of collisions
of the gas molecules against the vanes will
cause the wheel to rotate by one notch in the allowed
direction, but presumably never in the forbidden
direction.
Such rectification of thermal noise could be harnessed
to perform useful work (such as lifting a flea
against gravity), in direct violation of the Second Law.
Of course, in order for statistical fluctuations to
cause rotation at a perceptible rate, the ratchet
and pawl must be microscopic, and this points to
the resolution of the paradox.
If thermal motions of the gas molecules are
sufficient to cause the wheel to rotate a notch,
then the thermal motion of the pawl itself will
occasionally cause it to disengage
from the ratchet, at which point the wheel could
move in the ``forbidden'' direction.
Feynman compared the rates of the two processes --
rotation in the allowed and the forbidden
directions -- and found them to be equal when
the system is maintained at a single temperature.
Thus no net rotation arises, and the Second Law is
saved.

Since the failure of the ratchet and pawl system
to perform work
arises from thermal fluctuations of the pawl, a
natural solution to the problem is to reduce these
fluctuations by externally cooling the pawl to a
temperature below that of the gas.
In this case the device does indeed operate as
designed, but this no longer constitutes a violation
of the Second Law:
the ratchet and pawl is now effectively a
microscopic heat engine, capitalizing on a
temperature difference to extract useful work
from thermal motions.

While the ratchet and pawl was introduced in
Feynman's {\it Lectures} primarily for pedagogical
purposes, recent years have seen a renewed interest
in this system\cite{a,b,c,d,e,f,g,h,i,j,k,l,m1,m2,m3,n,o,p,q}, 
largely due to the fact that analogous mechanisms have
been proposed as simple models of motor proteins.

Our purpose in this paper is to introduce
an exactly solvable model of Feynman's microscopic
heat engine.
This model is discrete rather than 
continuous\footnote{
Note that a particle (or, more generally, a reaction
coordinate), evolving from one sufficienctly deep potential
(or free) energy minimum to another, behaves much as
if hopping from one site to another on a discrete
lattice.  See, for instance, Fig.6 of Ref.\cite{l}.},
but it captures two essential features of the
original example:
(1) a periodic but asymmetric interaction potential
between the ratchet and the pawl (corresponding
to the sawtooth shape for the ratchet's teeth), and
(2) two ``modes'' of interaction (corresponding
to the pawl being either engaged or disengaged
from the ratchet).
These features are sufficient for the model to reproduce 
the behavior discussed by Feynman.
Related discrete models of noise-induced transport
have appeared in the literature\cite{m1,m2,m3};
however, we believe ours to be the first exactly solvable
model in which (as in Feynman's example) the transport 
is explicitly driven by a temperature difference between 
two reservoirs.

In Section \ref{sec:noload}, we will introduce
our model -- a system in contact with two heat
reservoirs -- and consider it in the absence of an
external load.
We will solve for the net rate at which
the device produces directed motion, in terms of
the reservoir temperatures $T_A$ and $T_B$ 
(Eq.\ref{eq:drift}).
We will also solve for the net rates of
heat flow from reservoir $A$ to the system,
and from the system to reservoir $B$;
see Fig.\ref{fig:qdots}.
For zero external load, these rates, $\dot Q_A$ and 
$\dot Q_B$, must be equal, as we will indeed find 
them to be.
We will finally use these results to solve for
the entropy production rate.

In Section \ref{sec:load}, we will allow the device 
to perform work against an external load, $f$.
We will again solve exactly for the directed
motion and heat flows,
in terms of ${\bf x}\equiv(f,T_A,T_B)$.
In this situation, the heat flows $\dot Q_A$ and 
$\dot Q_B$ are not necessarily equal; the difference
between them is the {\it power}, $\dot W$,
which the device delivers against the load $f$.
When $\dot W>0$, the system operates as a heat
engine.
Conversely, one can imagine that for some 
parameter values, the system will operate as a
refrigerator, creating a net flow of heat from
the colder to the hotter reservoir.
In Section \ref{sec:eng_ref}
we will use our analytical results to show that
our model indeed exhibits both these
behaviors (heat engine and refrigerator).
Finally, in Section \ref{sec:linresp} we will
consider the near-equilibrium regime of small load 
and temperature difference.
We will show that our model confirms
general predictions based on linear response theory.

\section{Zero external load}
\label{sec:noload}

Consider a particle which jumps
between neighboring sites along a one-dimensional regular
lattice, where $d$ is the lattice spacing.
We assume that the particle is coupled to a heat
reservoir at temperature $T_B$, and that its
jumps are thermal in nature.
That is, the probability (per unit time) of making a
jump to site $i+1$, starting from site $i$,
is related to the probability rate of the inverse process
by the usual detailed balance relation:
$P_{i\rightarrow i+1}/P_{i+1\rightarrow i} =
\exp(-\Delta E/T_B)$,
where $\Delta E = U_{i+1}^{(m)}-U_i^{(m)}$ is the
instantaneous change in the particle's potential energy,
associated with the jump from $i$ to $i+1$.
We use the notation $U_i^{(m)}$ to denote the
potential energy of the particle at site $i$;
the superscript $m$ denotes the ``mode'' of the
potential, to be explained momentarily.
The integer index $i$ runs from $-\infty$ to
$+\infty$, and we are using units in which Boltzmann's
constant $k_B=1$.

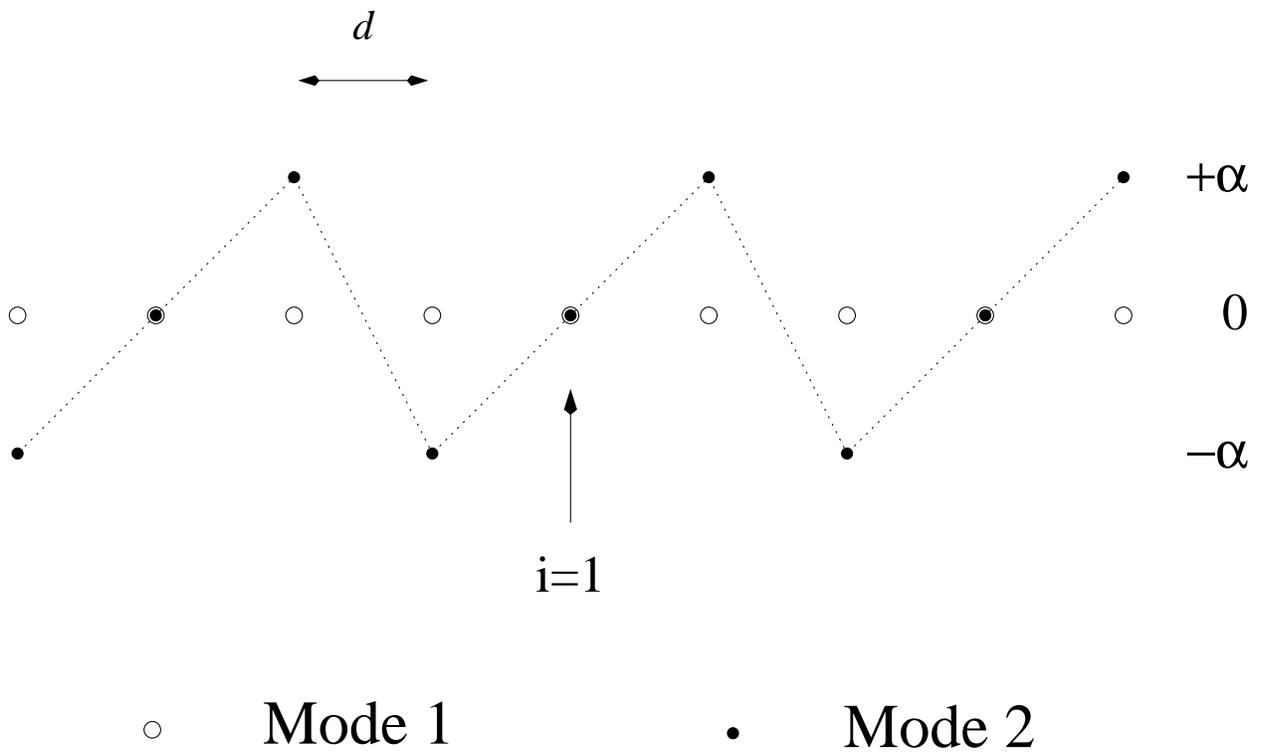
\begin{figure}
\caption{The potential energy $U_i^{(m)}$ is shown for
both modes, $m=1,2$, in the absence of an external load;
the lattice spacing is $d$, and site 1 is labelled explicitly.}
\label{fig:pot}
\end{figure}

Next, we assume that the potential energy function
$U_i^{(m)}$ has two possible {\it modes}, $m=1,2$,
and that it changes stochastically between these two.
In the first mode, the energy of each
site is zero: $U_i^{(1)}=0$.
In the second mode, the energy is periodic:
$U_i^{(2)}= \alpha \cdot [(i~{\rm mod}~3) - 1]$,
where $\alpha$ is a positive constant with units
of energy.
As shown in Fig.\ref{fig:pot}, the second mode is a discrete
version of an asymmetric sawtooth potential.
We assume that the stochastic process governing
the changes between modes is also a thermal
process, driven by a heat reservoir at temperature
$T_A$.
Thus, the probability rate of a change to
mode 2, starting from mode 1, relative to the probability
rate of the reverse
mode change, is given by the detailed balance
factor $\exp(-\Delta E/T_A)$, where
$\Delta E = U_i^{(2)} - U_i^{(1)}$
is now the (site-dependent) change in particle energy
due to an instantaneous change from mode 1 to mode 2.

We now describe more precisely these two stochastic
processes, governing the jumps of the
particle, and the change between modes.
We assume the processes are independent,
and each is a Poisson process
occurring at a rate $\Gamma$.
That is, during every infinitesimal time interval
$\delta t$, there is a probability $\Gamma\delta t$
that the particle will attempt a jump to a neighboring
site.
An ``attempt'' looks as follows.
First, the particle decides (randomly, with equal
probability) whether to try jumping to the left
($-d$) or to the right ($+d$).
Then the Metropolis algorithm\cite{metro} is used
to satisfy detailed balance:
if the value of $\Delta E$ associated with the
jump is zero or negative, then the jump takes place;
if $\Delta E>0$, then the jump occurs with
probability $\exp(-\Delta E/T_B)$.
Similarly, during every infinitesimal time interval
$\delta t$, there is a probability $\Gamma\delta t$
that the mode will attempt to change, and
the attempt is accepted or rejected
according to the Metropolis algorithm (at temperature
$T_A$).

We have introduced three parameters which we will
view as being ``internal'' to the system:
$d$, $\alpha$, and $\Gamma$;
these essentially set the relevant length ($d$),
energy ($\alpha$), and time ($\Gamma^{-1}$) scales.
The two remaining parameters
$T_A$ and $T_B$, we will view as ``external''.

The analogy between our model and Feynman's
ratchet and pawl runs as follows.
First, the position of our particle corresponds
to an angle variable, $\theta$,  specifying
the orientation of the ratchet wheel.
When the particle accomplishes a net displacement
of three lattice sites, to the right or
to the left, this is equivalent to the
ratchet being displaced by one notch, or
tooth\footnote{
Thus, the distance $3d$ in our model translates
to an angular interval $\Delta\theta=2\pi/N_{teeth}$,
where $N_{teeth}$ is the number of teeth along
the perimeter of the ratchet wheel.}; see
Fig.\ref{fig:pot}.
Since we want to keep track of the wheel over long
intervals of time, possibly including many full
rotations, $\theta$ varies from $-\infty$
to $+\infty$, rather than being a periodic
variable from 0 to $2\pi$.

When a ratchet and pawl are ``engaged''
-- that is, when the pawl actually presses
against the teeth of the ratchet --
then there exists an interaction energy,
arising from the compression of the spring
which holds the pawl against the ratchet,
which has the form of a periodic sawtooth
potential in the variable $\theta$.
In our model, the analogue of this
interaction energy is the discrete sawtooth
potential $U_i^{(2)}$;
mode 2 thus corresponds to the situation in
which the ratchet presses against the pawl.
By contrast, mode 1 corresponds to the ratchet
and pawl begin ``disengaged'', as will occur
every so often as a result of thermal fluctuations
of the pawl.
Of course, in Feynman's system,
the potential energy of the disengaged mode is
always greater than that of the engaged mode
(due to the spring compression needed to actually
place the pawl out of reach of the ratchet's teeth),
whereas in our model this is not the case:
$U_i^{(1)}=0$.
This, however, does not change
the problem in any qualitatively significant way.

As mentioned, the motion of the particle from
site to site is a thermal process occurring at
temperature $T_B$, while
the stochastic ``flashing'' between modes occurs
at temperature $T_A$.
Thus, in the context of the analogy with the
physical
ratchet and pawl, $T_B$ denotes the temperature
of the gas surrounding the panes connected
to the ratchet wheel, and $T_A$ is the temperature
at which the pawl is maintained.

To analyse the model, we first note that it maps
nicely onto the problem of 
a spin-1/2 particle, $A$, coupled to
a spin-1 particle, $B$, with the following
energy function:
\begin{equation}
E(S_A,S_B) = \alpha\cdot \biggl(S_A+{1\over 2}\biggr)
\cdot S_B,
\end{equation}
where $S_A=\pm 1/2$ and $S_B=0,\pm 1$.
Here, spin $A$ represents the mode (pawl),
spin $B$ represents the particle (ratchet).
When spin $A$ is ``down'' ($S_A=-1/2$), the energy
of the system is independent of the state of
spin $B$, as with mode 1;
when spin $A$ is ``up'', the energy is $-\alpha$,
0, or $+\alpha$, depending on the state of spin $B$,
as with mode 2.
Thus, changes in the state of spin $A$ correspond
to changes in the mode, whereas changes in the
state of spin $B$ correspond to the particle
making a jump.
If $S_B$ changes from
-1 to 0, or from 0 to +1, or from +1 to -1,
then this amounts to
the particle jumping to the right;
the reverse processes (0 to -1, etc.)
correspond to jumps to the left.
We now couple spin $A$ to a reservoir at temperature
$T_A$, spin $B$ to a reservoir at temperature $T_B$,
with the stochastic dynamics as outlined above (two
independent Poisson processes),
and solve for the net drift of the particle.

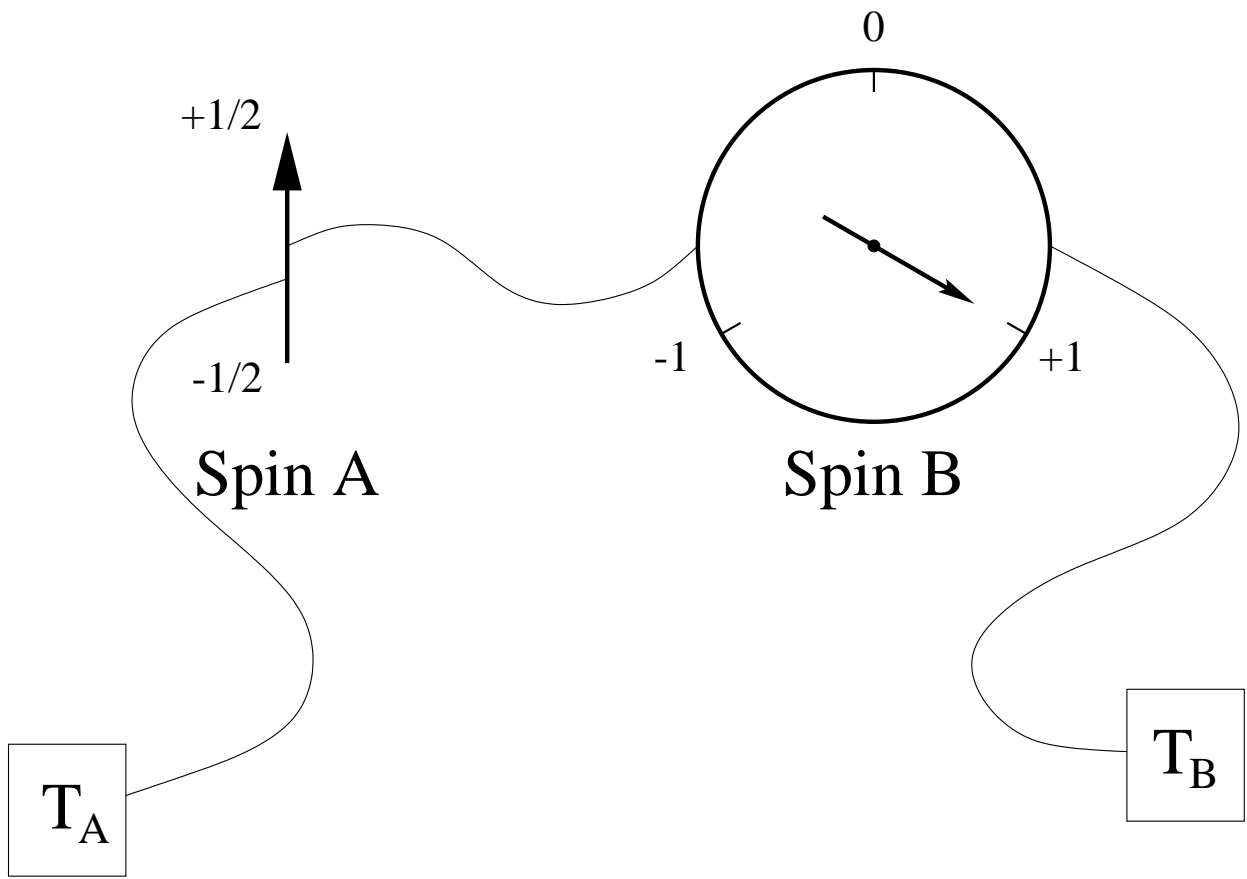
\begin{figure}
\caption{Our system maps onto that of a spin-1/2
particle ($A$) coupled to a spin-1 particle ($B$).
The former is depicted in the usual manner
(as ``up'' or ``down''),
the latter by an arrow which can point
in one of three directions on the face of a clock.
The thin lines denote the coupling
between the two spins, as well as the coupling
of each spin to a heat reservoir.}
\label{fig:spins}
\end{figure}

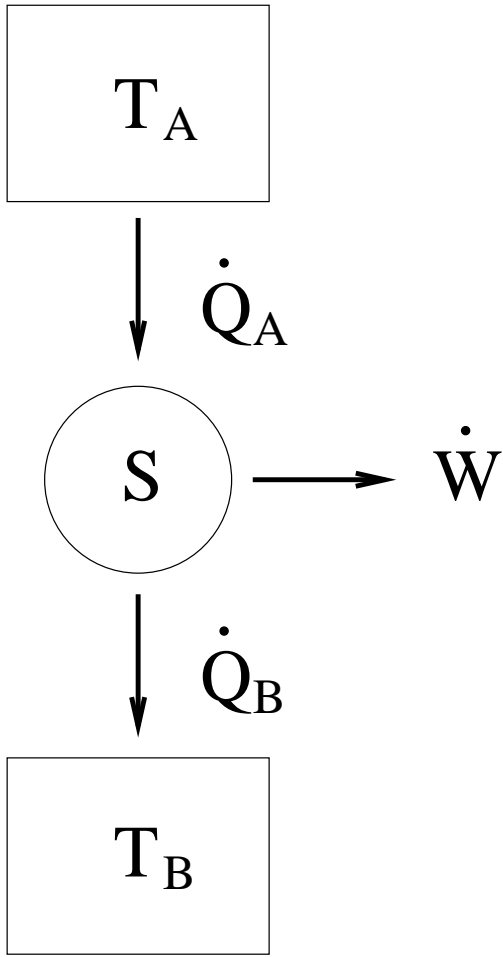
\begin{figure}
\caption{A schematic representation of our system
($S$) in contact with two heat reservoirs
(at temperatures $T_A$ and $T_B$).
As implied by the arrows, we define
$\dot Q_A$ to be the net flow of heat from
reservoir $A$ to the system, and $\dot Q_B$
to be the heat flow from the system to
reservoir $B$.
Therefore $\dot W=\dot Q_A - \dot Q_B$ is the
power delivered as work against an external
load.
(In Section \ref{sec:noload}, there is no such
load, hence $\dot Q_A=\dot Q_B$.)}
\label{fig:qdots}
\end{figure}

This system can be visualized as shown in Fig.\ref{fig:spins}.
Spin $A$ flips between up and down; spin $B$
performs sudden ``rotations'' by $\pm 120^\circ$.
A clockwise rotation corresponds to a
jump to the right ($+d$) by the particle;
counterclockwise ones translate into jumps to
the left.

\vspace{.2in}

\noindent
Table 1.~~~~~~~~~~~~~
\begin{tabular}{c||c|c|c}
~State (n)~~ & ~~~$S_A$~~~ & ~~~~$S_B$~~~~ & $E(S_A,S_B)$ \\ \hline
1     & -1/2  & -1    & 0            \\
2     & -1/2  & 0     & 0            \\
3     & -1/2  & +1    & 0            \\
4     & +1/2  & -1    & $-\alpha$    \\
5     & +1/2  & 0     & 0            \\
6     & +1/2  & +1    & $+\alpha$    \\
\end{tabular}

\vspace{.2in}

Our system has six possible
states, listed in Table 1.
The dynamics of the particle is described by
a Markov process, for which we can write
a set of rate equations.
Let $p_n(t)$ denote the probability that the
system is found in state $n$ at time $t$.
Then 
\begin{equation}
\label{eq:rates}
{dp_n\over dt} =
\Gamma\sum_{n^\prime\ne n}
\biggl[
p_{n^\prime} P(n^\prime\rightarrow n) -
p_n P(n\rightarrow n^\prime)
\biggr]\qquad,\qquad
n=1,\cdots,6.
\end{equation}
Here, $\Gamma\cdot P(n\rightarrow n^\prime)$ is the
probability rate at which the system, when its
state is $n$, makes a transition to state $n^\prime$.
If the transition $n\rightarrow n^\prime$
involves only a flip of spin $A$, 
then $P(n\rightarrow n^\prime)$ is the probability
that such an attempt will be accepted, under the
Metropolis rule.
If the transition involves just a rotation of spin $B$
(by $\pm 120^\circ$),
then $P(n\rightarrow n^\prime)$ is the probability
of generating this move, given an attempt to change
$S_B$.
If the transition involves changes in the states
of both spins, then
$P(n\rightarrow n^\prime)=0$.
Examples of these rules are:
\begin{equation}
\begin{array}{lll}
P(1\rightarrow 4)=1 \qquad&
P(1\rightarrow 2)={1\over 2} \qquad&
P(1\rightarrow 5)=0 \\
P(3\rightarrow 6)=e^{-\alpha/T_A} \qquad&
P(4\rightarrow 5)={1\over 2}\,e^{-\alpha/T_B} \qquad&
P(4\rightarrow 2)=0.
\end{array}
\end{equation}
The factors of one-half come from the fact that,
when the system attempts to
rotate spin $B$, there are two possible states for
which it can aim.

Our rate equations can be expressed as:
$d{\bf p}/dt=\Gamma{\cal R}{\bf p}$,
where ${\bf p}=(p_1,\cdots,p_6)^T$,
\begin{equation}
{\cal R} ~~=~~
\left(
\begin{array}{cccccc}
     -2~~~~~  & {1\over 2}~~~~ & {1\over 2}      & \mu & 0 & ~~~0 \\
     {1\over 2}~~~~~ & -2~~~~  & {1\over 2}      & 0   & 1 & ~~~0 \\
     {1\over 2}~~~~~ & {1\over 2}~~~~ & -1-\mu & 0   & 0 & ~~~1 \\
     1~~~~~   & 0~~~~   & 0        & -\mu-{\nu+\nu^2\over 2}
                          & {1\over 2} & ~~~{1\over 2} \\
     0~~~~~   & 1~~~~   & 0        & {\nu\over 2}  & -{3+\nu\over 2}
                                       & ~~~{1\over 2} \\
     0~~~~~   & 0~~~~   & \mu      & {\nu^2\over 2} & {\nu\over 2}
              & ~~~-2
\end{array}
\right),
\end{equation}
and we have introduced the constants
\begin{equation}
\mu = e^{-\alpha/T_A} \qquad{\rm and}\qquad
\nu = e^{-\alpha/T_B}.
\end{equation}
Note that $\mu$ and $\nu$ are monotonically
increasing functions of $T_A$ and $T_B$, and
the temperature range $0<T_A~(T_B)<\infty$
translates to $0<\mu~(\nu)<1$.
We thus think of $\mu$ and $\nu$ is ``rescaled''
temperatures.

The long-time behavior of our system is governed
by the steady-state distribution of probabilities,
$\bar{\bf p}$, which is the null
eigenvector of ${\cal R}$
(i.e.\ ${\cal R}\bar{\bf p} = 0$).
Determining $\bar{\bf p}$ is an exercise in Jordan 
elimination, and leads to the following result:
$\bar{\bf p}={\bf x}/N$, where
\begin{eqnarray}
x_1 &=& 52\mu + 28\mu^2 + 12\nu + 19\nu^2 + 5\nu^3 +
        21\mu\nu + 2\mu\nu^2 + 8\mu^2\nu \\
x_2 &=& 36\mu + 16\mu^2 + 28\nu + 27\nu^2 + 5\nu^3 +
        25\mu\nu + 8\mu\nu^2 + 2\mu^2\nu \\
x_3 &=& 44\mu + 19\mu\nu + 20\nu + 49\nu^2 + 15\nu^3 \\
x_4 &=& 64 + 20\nu + 48\mu + 15\mu\nu \\
x_5 &=& 24\mu + 40\nu + 20\nu^2 + 18\mu^2 + 30\mu\nu +
        15\mu\nu^2 \\
x_6 &=& 22\mu^2 + 16\mu\nu + 44\mu\nu^2 + 14\mu^2\nu +
        15\mu\nu^3 + 26\nu^2 + 10\nu^3,
\end{eqnarray}
and $N(\mu,\nu) = \sum_{i=1}^6 x_i$ is a normalization
factor.

When both temperatures go
to zero, $\mu,\nu\rightarrow 0$, we get
$\bar{\bf p}^T = (0,0,0,1,0,0)$.
This makes sense: in that limit, the system freezes
to the state of lowest energy, $n=4$.

We now address the question of net drift.
We first define a net current, $J\equiv J_+ - J_-$,
where $J_+$ is the rate at which spin $B$ is
observed to change from state 0 to state +1,
and $J_-$ is the rate of the reverse transitions,
from +1 to 0.
This can be interpreted by imagining an observer
placed at 2 o'clock on the clock face depicting spin $B$
in Fig.\ref{fig:spins};
$J_+$ ($J_-$) is then the rate at which the hand of
the clock passes that observer in the clockwise
(counterclockwise) direction;
by ``rate'', we mean number of passes
per unit time, averaged over an infinitely long
interval of time.\footnote{
Of course, we could just as well have placed our
observer at 6 or 10 o'clock;
in the steady state, the current measured will be
independent of where the observer is placed.}
The current $J$ represents the {\it net} average
rate of clockwise revolutions of spin $B$.
Since each revolution
corresponds to 3 steps of the particle to the right,
this translates into a particle drift:
\begin{equation}
\label{eq:vj}
v = 3\,dJ,
\end{equation}
where $v$ denotes the (steady-state) average velocity
of the particle.

Explicit expression for the quantities $J_\pm$ are given by:
\begin{mathletters}
\label{eq:jpm}
\begin{eqnarray}
J_+ &=& \Gamma \, (\bar p_2 {\cal R}_{32} + \bar p_5 {\cal R}_{65}) \\
J_- &=& \Gamma \, (\bar p_3 {\cal R}_{23} + \bar p_6 {\cal R}_{56}).
\end{eqnarray}
\end{mathletters}
(Since $\Gamma{\cal R}_{mn}$ is the transition rate
to state $m$, {\it given} that the system is found in state $n$,
$\bar p_n\Gamma{\cal R}_{mn}$ is the net rate at
which transitions from $n$ to $m$ are observed to occur in
the steady state.)
Using our results for $\bar{\bf p}$, we 
get, after some algebra:
\begin{equation}
\label{eq:drift}
v(T_A,T_B) = -3\, d\,{\Gamma\over N} (\mu-\nu) (1-\nu) (3\mu+4).
\end{equation}

There are a number of things to note about this result.
First, it implies that if $T_A>T_B$ (i.e.\ $\mu>\nu$),
then there is a net flow of the particle to the left;
if $T_B>T_A$, the particle drifts to the right.
If $T_A=T_B$, then there is no drift,
in agreement with Feynman's analysis (as well as the
Second Law!).

Next, notice that $v\rightarrow 0$ as
$T_B\rightarrow\infty$ ($\nu\rightarrow 1$).
In that limit, the change in energy arising from a
jump to the left or to the right becomes negligible
in comparison to the temperature of the reservoir
which drives those jumps;
thus, from any lattice site, the particle is as likely
to jump to the left as to the right, resulting in no
net drift.

Finally, in the limit $T_A\rightarrow\infty$
($\mu\rightarrow 1$), we get:
\begin{equation}
v = -21\, d\,{\Gamma\over N} (1-\nu)^2
\qquad,\qquad
T_A\rightarrow\infty.
\end{equation}
This is the limit in which changes between the modes
occur independently of location of the particle:
every attempt to change the mode is accepted.
This is analogous to the situation studied by
Astumian and Bier\cite{b}, where the ``flashing'' between
the two modes of the potential is a Poisson process
independent of the particle position.

We can also compute the average rates at
which heat is transferred between the two reservoirs
and our system.
Whenever the system makes a transition from state 1
to state 4, or from state 6 to state 3, its energy drops
by $\alpha$; this energy is released into the reservoir
at temperature $T_A$.
Conversely, during the transitions $4\rightarrow 1$
and $3\rightarrow 6$, the system absorbs energy
$\alpha$ from reservoir $A$.
The net rate at which the system absorbs
energy from reservoir $A$ is then:
\begin{equation}
\label{eq:qadot}
\dot Q_A =
\alpha\Gamma\,
\Bigl(\bar p_4{\cal R}_{14} + \bar p_3{\cal R}_{63}
-\bar p_1{\cal R}_{41} - \bar p_6{\cal R}_{36}\Bigr).
\end{equation}
We can similarly write down an expression for
the rate at which heat flows from our
system to reservoir $B$:
\begin{equation}
\label{eq:qbdot}
\dot Q_B =
\alpha\Gamma\,
\Bigl(\bar p_5 {\cal R}_{45} +
\bar p_6 {\cal R}_{56} +
2 \bar p_6 {\cal R}_{46} -
\bar p_4 {\cal R}_{54} -
\bar p_5 {\cal R}_{65} -
2 \bar p_4 {\cal R}_{64}).
\end{equation}
Fig.\ref{fig:qdots} illustrates the sign convention
which we choose in defining $\dot Q_A$ and $\dot Q_B$.
Plugging in the values for the components
of $\bar{\bf p}$ and ${\cal R}$, we find that
$\dot Q_A = \dot Q_B$, as we could have predicted,
since in the steady state there is no net absorption
of heat by the system, nor is any of the heat delivered
as work against an external load.
Thus, the particle drift is driven by a net
passage of heat from $A$ to $B$, by way of the system.
The explicit expression for this heat flow is:
\begin{equation}
\label{eq:heatflow}
\dot Q_{A\rightarrow B} =
\dot Q_A = \dot Q_B =
3 \, {\alpha\Gamma\over N}
(\mu-\nu) \,P(\mu,\nu),
\end{equation}
where
$P(\mu,\nu) = 4 +14\mu +15\nu +4\mu\nu +5\nu^2 > 0$.
The ratio $v/\dot Q_{A\rightarrow B}$ then gives us
the average displacement of the particle, per unit
of heat passed (via the system) from reservoir $A$ to
reservoir $B$:
\begin{equation}
\label{eq:dxdQ}
\lim_{\tau\rightarrow\infty}
{\Delta x\over\Delta Q} =
{v\over\dot Q_{A\rightarrow B}} =
- {d\over\alpha}\cdot
{(1-\nu)(3\mu+4)\over P(\mu,\nu)} < 0.
\end{equation}
Here, $\Delta x$ and $\Delta Q$ are the net particle
displacement, and the net heat transferred from
$A$ to $B$, over a time interval $\tau$.
The factor $(\mu-\nu)$ in Eq.\ref{eq:heatflow} guarantees
that the direction of the heat flow is from the hotter
to the cooler reservoir.

We can also compute the rate at which entropy
is produced during this process.
The rate of entropy production associated with the
flow of heat from reservoir $A$ to the system is:
$\dot S_A = -\dot Q_A/T_A$; and for reservoir $B$:
$\dot S_B = \dot Q_B/T_B$.
The net entropy production rate is thus:
\begin{eqnarray}
\dot S &=& \dot S_A + \dot S_B
= {T_A-T_B\over T_AT_B}\,\dot Q_{A\rightarrow B} \\
&=& 3 {\Gamma\over N} \biggl(\ln{\mu\over\nu}\biggr)
\,(\mu-\nu) \,P(\mu,\nu)\ge 0.
\end{eqnarray}

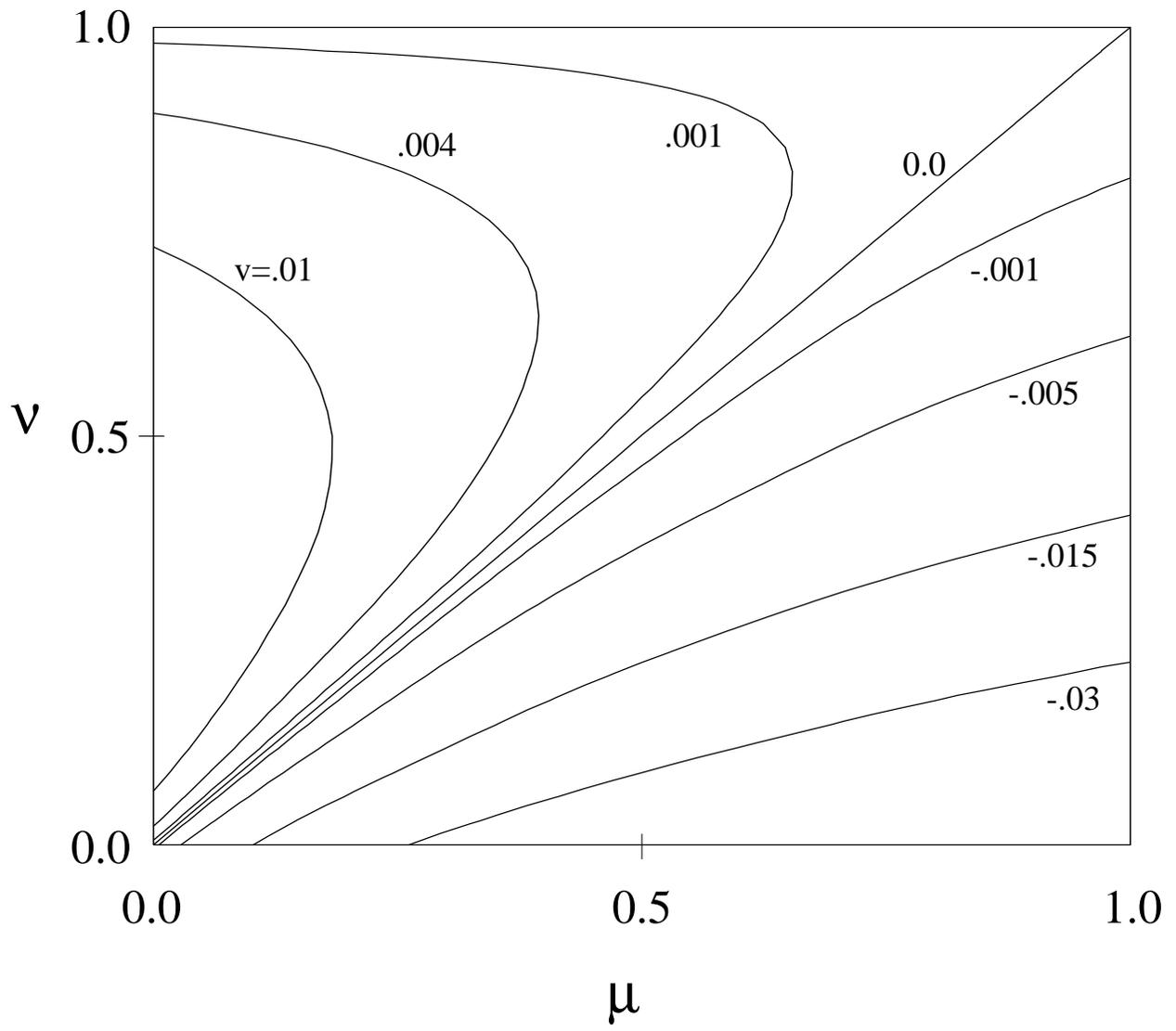
\begin{figure}
\caption{A contour plot of the function
$v(\mu,\nu)$, where
$\mu = \exp(-\alpha/T_A)$ and
$\nu = \exp(-\alpha/T_B)$ are the rescaled
temperatures of the two reservoirs.}
\label{fig:contour}
\end{figure}

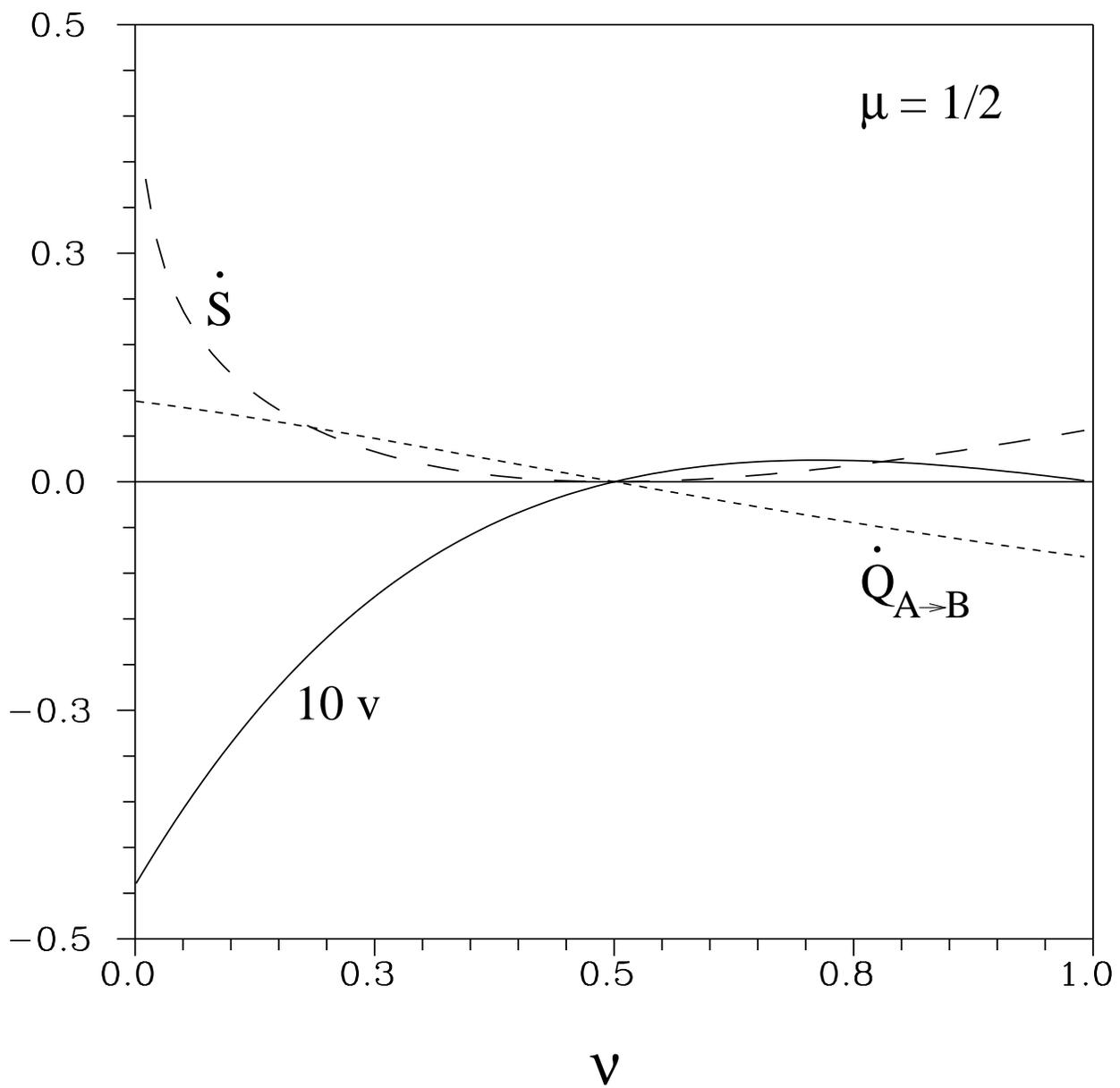
\begin{figure}
\caption{The drift $v$ (multiplied by 10), heat flow
$\dot Q_{A\rightarrow B}$, and rate of entropy production
$\dot S$, plotted as functions
of $\nu$, for fixed $\mu=1/2$.}
\label{fig:noload}
\end{figure}

Figs.\ref{fig:contour} and \ref{fig:noload} illustrate
the results obtained in this section.
Fig.\ref{fig:contour} is a contour plot of the drift
$v$, as a function of the rescaled reservoir
temperatures $\mu$ and $\nu$.
The contour $v=0$ runs along the
diagonal, $\mu=\nu$, as well as along
$\nu=1$ ($T_B\rightarrow\infty$).
The appearance of positive contours ($v>0$) to the
left of the diagonal, and negative ones to the
right, illustrates the point that the drift is
rightward when $T_B>T_A$ and leftward when $T_A>T_B$.

In Fig.\ref{fig:noload} we have fixed the value of
$T_A$ by setting $\mu=1/2$, and have plotted $v$,
$\dot Q_{A\rightarrow B}$, and $\dot S$ as functions
of $\nu$.
All three quantities hit zero at $\nu=1/2$, where
$T_A=T_B$: nothing interesting happens when the
system is maintained at a single temperature.
Note also that $v$ and $\dot Q_{A\rightarrow B}$
are opposite in sign (in agreement with Eq.\ref{eq:dxdQ}),
while $\dot S$ is always nonnegative (in agreement with
the Second Law).
Finally, note that $v\rightarrow 0$ as $\nu\rightarrow 1$
($T_B\rightarrow\infty$).

In plotting these two figures, we set all the internal
parameters to unity: $\alpha=d=\Gamma=1$.

\section{Non-zero external load}
\label{sec:load}

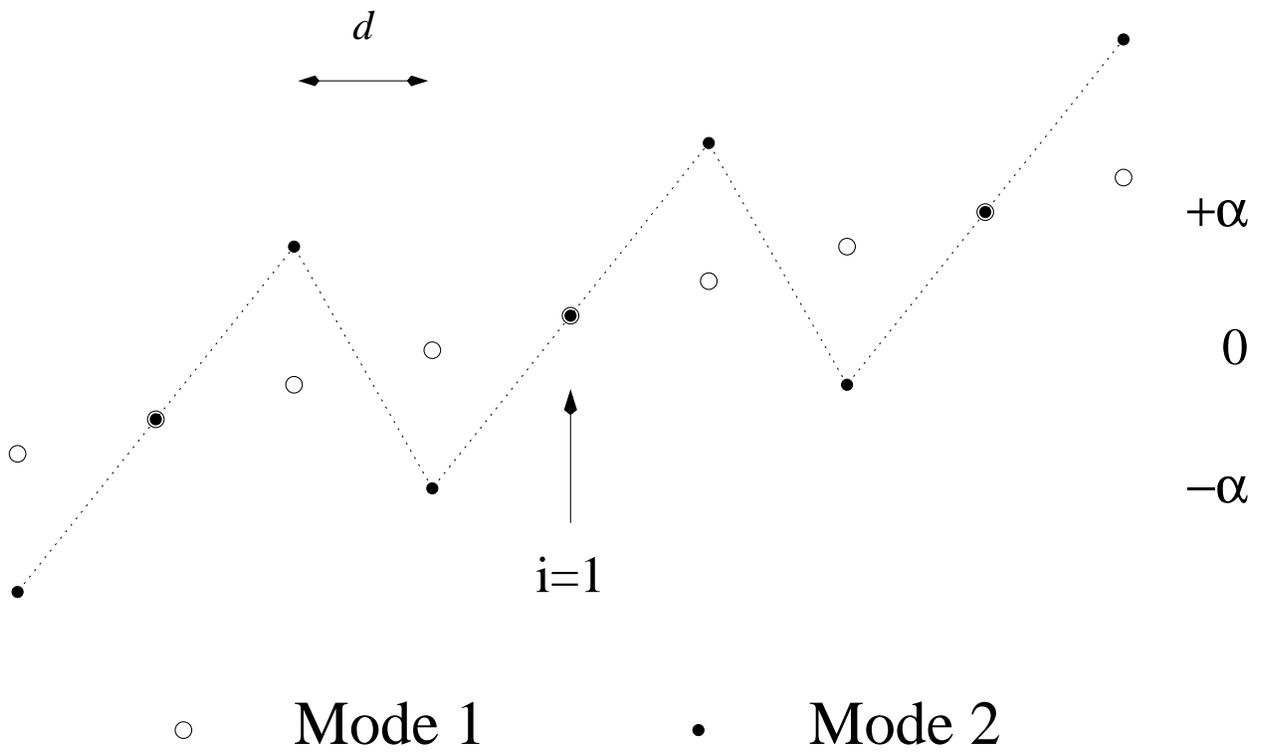
\begin{figure}
\caption{The discrete, two-mode potential energy
function $U_i^{(m)}$ is illustrated for the
situation in which there is a non-zero external load.
Here we have chosen $f=\alpha/4d$.}
\label{fig:load}
\end{figure}

In this section we add an external load to our model.
In Feynman's example, this load is a flea,
attached by a thread to the ratchet wheel:
when the wheel rotates in the appropriate direction,
the creature is lifted against gravity.
In our model, we add a slope to the discrete
potential:
\begin{equation}
U_i^{(m)} \rightarrow U_i^{(m)} + ifd ,
\end{equation}
where $f$ is a real constant (and $i$ is the
lattice site); see Fig.\ref{fig:load}.
Effectively, $f$ is a constant external force
which pulls the particle leftward if $f>0$,
rightward if $f<0$.

The presence of an external load allows the system
to perform {\it work}.
If, in the steady state achieved for fixed
${\bf x}=(f,T_A,T_B)$, the particle experiences
a drift $v({\bf x})$, then the {\it power}
delivered against the external load is:
\begin{equation}
\label{eq:wdot}
\dot W({\bf x}) = f v({\bf x}).
\end{equation}
By conservation of energy, this must be balanced by
heat lost by the reservoirs:
\begin{equation}
\label{eq:econs}
\dot W = \dot Q_A - \dot Q_B.
\end{equation}

Our approach to solving for the steady-state behavior
is the same as in Section \ref{sec:noload}, except
that the presence of the term $ifd$ in the potential
changes the elements of ${\cal R}$, and therefore the
steady-state probabilities $\bar{\bf p}$.

Because the acceptance probability of an attempted
move in the Metropolis scheme has the form
${\rm Prob.} = {\rm min}\{1,\exp-\Delta E/T\}$,
each matrix element ${\cal R}_{ij}$ is
piecewise analytic in $f$. 
A little thought reveals that the
$f$-axis can be divided into four ranges of values,
over each of which the elements of ${\cal R}$ can be
written as analytic functions of $\alpha$, $d$, $f$,
$T_A$, and $T_B$.
These ranges are:
\begin{mathletters}
\begin{eqnarray}
-\infty < &f& < -\alpha/d \\
-\alpha/d < &f& < 0 \\
0 < &f& < 2\alpha/d \\
2\alpha/d < &f& < +\infty.
\end{eqnarray}
\end{mathletters}
In the two extreme ranges, (a) and (d), stretching
to $-\infty$ and $+\infty$, the slope is so steep
that the potential energy function no longer has a
sawtooth shape in mode 2.
We will ignore these ranges and focus instead
on the ones for which $U_i^{(2)}$ {\it does} look
like a discrete sawtooth.

For range (b), i.e.\ $-\alpha/d < f < 0$,
the potential slopes downward with increasing $i$,
and an explicit expression for ${\cal R}$ is:
\begin{equation}
{\cal R}^b ~~=~~
\left(
\begin{array}{cccccc}
     -{3\over 2} - {1\over 2\sigma} & {1\over 2\sigma} & {1\over 2}
           & \mu & 0 & ~~~0 \\
     {1\over 2} & -{3\over 2} - {1\over 2\sigma} & {1\over 2\sigma}
           & 0 & 1 & ~~~0 \\
     {1\over 2\sigma} & {1\over 2} & -{1\over 2}-\mu-{1\over 2\sigma}
           & 0 & 0 & ~~~1 \\
     1 & 0 & 0 & -\mu-{\nu^2\over 2\sigma}-{\nu\sigma\over 2}
           & {1\over 2} & ~~~{1\over 2} \\
     0 & 1 & 0 & {\nu\sigma\over 2} & -{3\over 2}-{\nu\sigma\over 2}
           & ~~~{1\over 2} \\
     0 & 0 & \mu & {\nu^2\over 2\sigma} & {\nu\sigma\over 2} & ~~~-2
\end{array}
\right),
\end{equation}
where
\begin{equation}
\sigma=e^{-fd/T_B}.
\end{equation}
For range (c), $0 \le f < 2\alpha/d$, the potential slopes
upward with $i$, and we have
\begin{equation}
{\cal R}^c ~~=~~
\left(
\begin{array}{cccccc}
     -{3\over 2} - {\sigma\over 2} & {1\over 2} & {\sigma\over 2}
           & \mu & 0 & ~~~0 \\
     {\sigma\over 2} & -{3\over 2} - {\sigma\over 2} & {1\over 2}
           & 0 & 1 & ~~~0 \\
     {1\over 2} & {\sigma\over 2} & -{1\over 2}-\mu-{\sigma\over 2}
           & 0 & 0 & ~~~1 \\
     1 & 0 & 0 & -\mu-{\nu^2\over 2\sigma}-{\nu\sigma\over 2}
           & {1\over 2} & ~~~{1\over 2} \\
     0 & 1 & 0 & {\nu\sigma\over 2} & -{3\over 2}-{\nu\sigma\over 2}
           & ~~~{1\over 2} \\
     0 & 0 & \mu & {\nu^2\over 2\sigma} & {\nu\sigma\over 2} & ~~~-2
\end{array}
\right).
\end{equation}
Note that both ${\cal R}^b$ and ${\cal R}^c$ reduce
to the matrix ${\cal R}$ of Section \ref{sec:noload}, when
$\sigma=1$, i.e.\ $f=0$.

As in Section \ref{sec:noload}, the first order of business
is to solve for 
$\bar{\bf p}$, using Jordan elimination, only now
the process is considerably more tedious:
terms do not cancel as nicely as when $f=0$.
The final results for the steady state probability
vectors $\bar{\bf p}^b$ and $\bar{\bf p}^c$
(corresponding to the two ranges of $f$ values) 
are of the form
\begin{equation}
\label{eq:pnj}
\bar p_n^j =
{ P_n^j(\mu,\nu,\sigma) \over
  N^j(\mu,\nu,\sigma) }
\qquad , \qquad j=b,c,
\qquad n=1\cdots 6,
\end{equation}
where the $P_n^j$'s are finite polynomials
in the variables $\mu$, $\nu$, and $\sigma$, and
$N^j=\sum_{n=1}^6 P_n^j$ is a normalization
factor.
Explicit expression for the polynomials
$P_n^j$ are given in the Appendix.

We can now obtain
$v({\bf x})$, $\dot Q_A({\bf x})$, and $\dot Q_B({\bf x})$
from $\bar{\bf p}$, as in Section \ref{sec:noload}.
The results for the two ranges of $f$ values,
$j=b,c$, are:
\begin{mathletters}
\label{eq:ss}
\begin{eqnarray}
\label{eq:ss1}
v^j({\bf x}) &=&
{3\over 2}d\,\Gamma\, {X^j\over\sigma N^j} \\
\dot Q_A^j({\bf x}) &=&
\alpha\Gamma\, {Y^j\over N^j} \\
\dot Q_B^j({\bf x}) &=&
\alpha\Gamma\, {Y ^j\over N^j} -
{3\over 2} fd\,\Gamma\, {X^j\over\sigma N^j} \qquad ,
\end{eqnarray}
\end{mathletters}
where $X^j(\mu,\nu,\sigma)$ and $Y^j(\mu,\nu,\sigma)$
are polynomials
for which explicit expressions are
presented in the Appendix.

Note that the steady-state behavior described by
Eq.\ref{eq:ss} clearly satisfies energy conservation
(see Eqs.\ref{eq:wdot} and \ref{eq:econs}):
\begin{equation}
\label{eq:dependent}
\dot Q_A - \dot Q_B = fv.
\end{equation}

Eq.\ref{eq:ss} is the central result of this paper
(and reduces to the results of Section \ref{sec:noload}
when $f=0$).
In Section \ref{sec:eng_ref}, we use Eq.\ref{eq:ss}
to show that our model can act both as a heat engine
and as a refrigerator.
In Section \ref{sec:linresp}, we consider the behavior
of our system near equilibrium, and we use Eq.\ref{eq:ss}
to confirm predictions based on a general, linear
response analysis.

\section{The system as heat engine and refrigerator}
\label{sec:eng_ref}

We can anticipate two different scenarios
in which our system acts as a ``useful'' device:

(1) $\dot W>0$.
In this case, the system is a {\it heat engine},
causing the particle to drift up the potential energy
slope, with efficiency
$\eta_{\rm eng} = \dot W/\dot Q_>$, where
$\dot Q_>$ is the rate of heat flow out of the hotter
reservoir.

(2) $\dot W<0$ and $\dot Q_<>0$, where $\dot Q_<$ is the
rate of heat flow out of the colder reservoir.
Here the system is a {\it refrigerator}, with
efficiency $\eta_{\rm ref} = \dot Q_</\vert\dot W\vert$.
The particle drifts down the potential slope,
and the resulting energy liberated allows for a net
transfer of heat from the colder to the hotter reservoir,
without violating the Second Law.

We will now use the results derived in
Section \ref{sec:load} to show that our simple model
indeed exhibits both these behaviors.

The system is a heat engine
when $v(f,T_A,T_B)$ and $f$ are of the same sign.
Let us introduce the variables
\begin{eqnarray}
\beta &=& {T_A+T_B \over 2T_AT_B} \\
\gamma &=& {T_A-T_B \over T_AT_B},
\end{eqnarray}
and consider the behavior of our system in
$(f,\gamma)$-space, for a fixed value of $\beta$.
(Note that $\beta$ is the
average inverse temperature of the reservoirs,
and $\gamma$ is the difference
between inverse temperatures.)
Because $\dot Q_A$, $\dot Q_B$, and $v$
are mutually dependent (Eq.\ref{eq:dependent}),
we can generically
explore a measurable fraction of the
space of steady-state behaviors by varying only two 
independent parameters, $f$ and $\gamma$,
while holding the third, $\beta$, fixed.

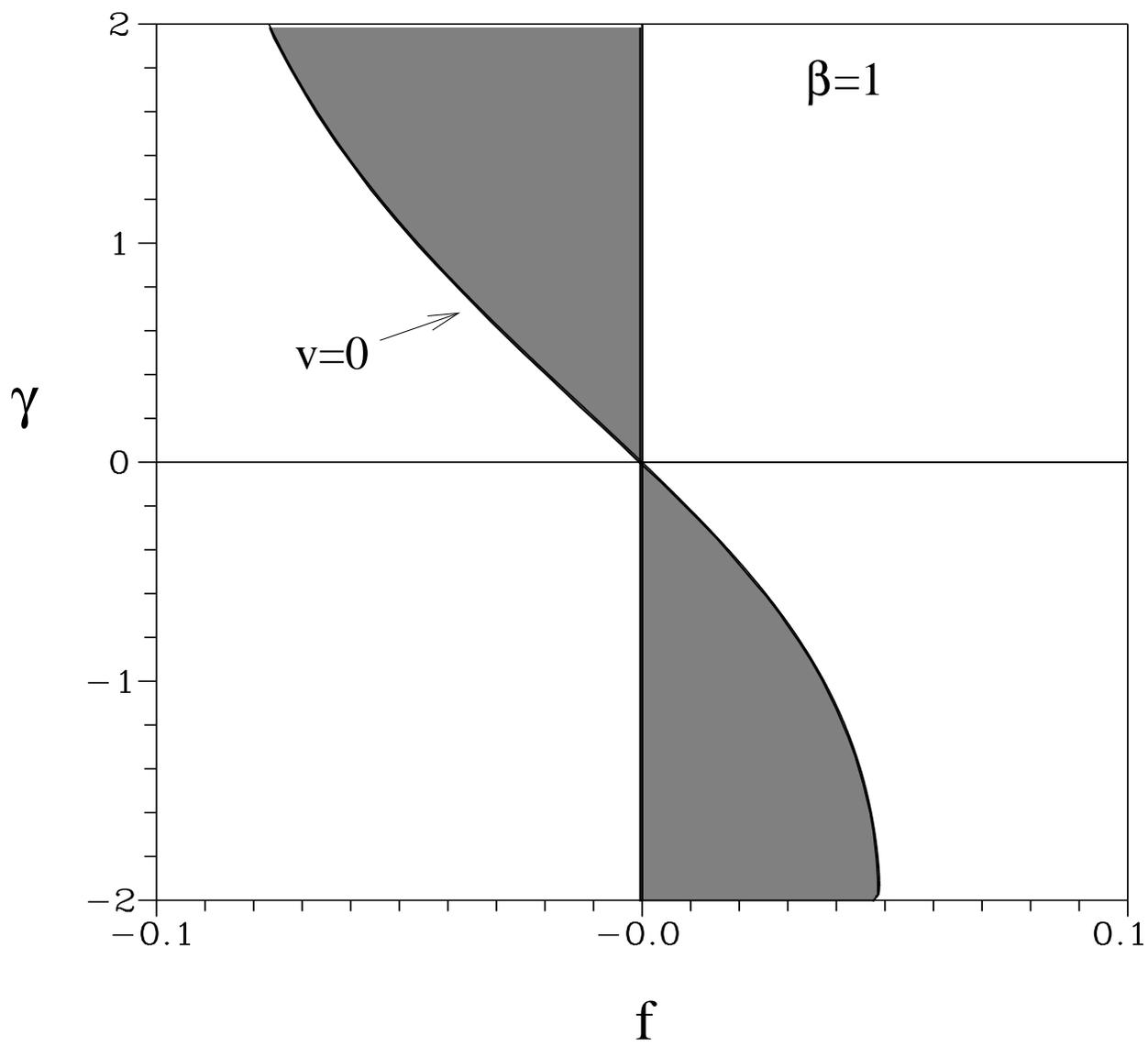
\begin{figure}
\caption{The contour $v(f,\gamma)=0$ is shown,
for fixed average inverse temperature $\beta=1$.
The shaded regions are those for which $fv>0$,
i.e.\ for which the system behaves as a heat
engine.}
\label{fig:engine}
\end{figure}

In Fig.\ref{fig:engine} we plot the contour
$v(f,\gamma)=0$, having set $\beta=1$, and
$\alpha=d=\Gamma=1$.
(The range of $\gamma$ values for this choice of
$\beta$ is $-2<\gamma<2$.)
To the left of this contour, we have $v>0$;
to the right, $v<0$.
The shaded region thus represents the values of
$(f,\gamma)$ for which $v$ and $f$ are of the same
sign, i.e.\ where the system behaves as a heat engine.

We can understand the placement of the shaded region as follows.
Consider a point $P$ on the positive $\gamma$-axis:
$f=0$, $\gamma>0$ (i.e.\ zero external load, $T_A>T_B$).
From Section \ref{sec:noload}, we know that the particle
then drifts leftward, $v<0$, although no work is performed,
since $f=0$.
Let us now imagine tilting the potential slightly
``downward'' ($f<0$).
For a small enough tilt, we expect that the particle will
continue to drift to the left, but now this drift is
uphill, and therefore work is done against the
external load: $\dot W>0$.
We conclude that points immediately to the left of the
positive vertical axis will correspond to external parameters
for which the system behaves as a heat engine, and
Fig.\ref{fig:engine} confirms this.
(Similar reasoning applies for the negative
$\gamma$-axis, where the region $\dot W>0$ appears to
the right.)
If we now continue to tilt the slope more and more
downward ($f$ increasingly negative), at fixed $\gamma>0$,
we expect the leftward drift to become progressively slower,
until for some tilt we get $v=0$.
At this point the leftward ``thermal force'' exerted on
the particle due to the temperature difference between
the reservoirs, exactly balances the external load.
This occurs at the boundary of the shaded region
(the contour $v=0$): for more negative slopes,
the particle slides down the slope, and the system
no longer acts as a heat engine.

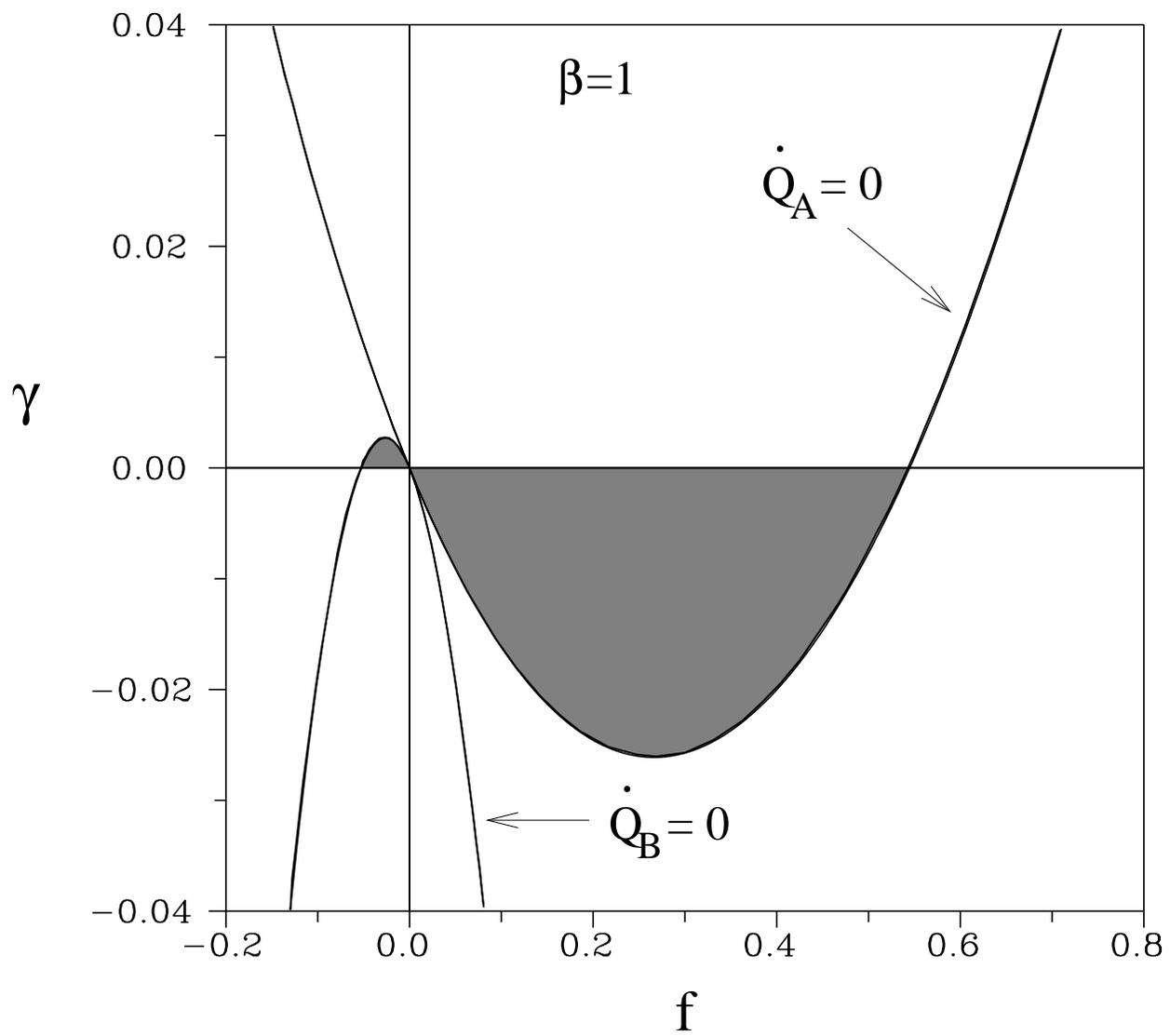
\begin{figure}
\caption{The contours $\dot Q_A(f,\gamma)=0$ and
$\dot Q_B(f,\gamma)=0$ are shown, for fixed $\beta=1$.
In the shaded regions, there is a net flow of heat
out of the colder reservoir; the system then acts as
a refrigerator.}
\label{fig:refrig}
\end{figure}

Our system is a refrigerator when
$\dot Q_<>0$, where $\dot Q_<$ is the rate at
which heat leaves the colder of the two reservoirs.
In Fig.\ref{fig:refrig} we plot the contours
$\dot Q_A(f,\gamma)=0$ and $\dot Q_B(f,\gamma)=0$,
again for $\beta=1$.
These two contours are tangent at the origin,
and divide the plane of $(f,\gamma)$-values as follows:
$\dot Q_A$ is positive for points lying above the contour,
and negative below.
Similary, $\dot Q_B>0$ ($\dot Q_B<0$) for points
lying above (below) the contour $\dot Q_B=0$.
Now, above the horizontal axis ($\gamma>0$) we have
$T_A>T_B$, hence $\dot Q_<=-\dot Q_B$.
The small shaded region in the second quadrant
therefore represents values of $(f,\gamma)$ for
which reservoir $B$ is the colder of the two
reservoirs, {\it and} it is losing heat ($\dot Q_B<0$).
Hence in this region our system acts as a refrigerator.
Below the horizontal axis, $T_B>T_A$ and thus
$\dot Q_<=\dot Q_A$.
The larger shaded region in the fourth quadrant thus
also represents refrigeration, only now reservoir
$A$ is the being drained of heat.

We can understand the general shape of the shaded
regions by assuming that, when the temperatures are
equal ($\gamma=0$) and the slope is very small,
the quantities $v$, $\dot Q_A$, and $\dot Q_B$
are linear functions of the slope:
\begin{mathletters}
\label{eq:ref_lin}
\begin{eqnarray}
v(f,0) &=& c_v f + O(f^2) \\
\dot Q_A(f,0) &=& c_A f + O(f^2) \\
\dot Q_B(f,0) &=& c_B f + O(f^2),
\end{eqnarray}
\end{mathletters}
with $c_v, c_A, c_B \ne 0$.
(Then $c_v<0$, since the particle
cannot slide {\it up} the slope when the reservoir
temperatures are equal.)
Energy conservation implies that the difference
between $\dot Q_A$ and $\dot Q_B$ must be quadratic in $f$
(since $\dot Q_A - \dot Q_B = \dot W = fv = c_v f^2$), hence
$c_A=c_B$.
Thus, for equal temperatures and a sufficiently small
slope, one of the reservoirs will be
losing heat and the other will be gaining it.
If we now slightly lower the temperature of the
reservoir which is losing heat, then we have a
refrigerator: heat flows out of the colder reservoir.
In our model, we have $c\equiv c_A=c_B>0$, hence for points on
the $\gamma=0$ axis immediately to the right of the origin,
heat flows out of reservoir $A$ and into reservoir $B$;
immediately to the left of the origin the reverse holds true.
This explains why, just to the right (left) of the origin,
the shaded region corresponding to refrigeration hugs the
horizontal axis from below (above).
If we continue to the right along the line $\gamma=0$,
increasing the value of $f$, then the particle will
drift ever more rapidly to the left as the slope becomes
ever more steeply inclined.
The potential energy lost as the particle slides down
the incline gets dissipated into the reservoirs;
for sufficiently large $f$, the rate of dissipation
is great enough that both reservoirs become heated:
$\dot Q_A<0$, $\dot Q_B>0$.
This happens to the right of the point
at which the contour $\dot Q_A=0$ crosses the
horizontal axis with a positive slope; for values of
$f$ beyond this point the system can no longer operate
as a refrigerator.

We can also understand why the two
contours $\dot Q_A=0$ and $\dot Q_B=0$ ``kiss'' at
the origin.
The result $c_A=c_B$ means that $\dot Q_A$ and
$\dot Q_B$ are (to leading order) equal along the
horizontal axis $\gamma=0$, near the origin.
However, they are also (exactly) equal along the
vertical axis, since $\dot W=0$ when $f=0$.
Thus, $\dot Q_A$ and $\dot Q_B$ are equal, to leading
order in $f$ and $\gamma$, for a small region around
the origin: $\dot Q_A = \dot Q_B = cf + b\gamma$.
This implies that their contours are both tangent
to the line $\gamma=-cf/b$ at the origin.

Since we have expressions for
$v({\bf x})$, $\dot Q_A({\bf x})$, and $\dot Q_B({\bf x})$,
we can compute the thermal efficiency
$\eta$, when it acts as either a heat engine or a
refrigerator.
By the Second Law, these efficiencies must never
exceed the Carnot efficiencies, $\eta_{\rm eng}^C$ and 
$\eta_{\rm ref}^C$ (which depend only on $T_A$ and $T_B$).
Ideally, we could use our exact results to find the
maximum {\it relative efficiency} ($y=\eta/\eta^C$) which 
our model achieves, both as a heat engine and
as a refrigerator.
Unfortunately, the expressions for $v$, $\dot Q_A$, and 
$\dot Q_B$ are sufficiently complicated
that we are not able to find these maxima analytically.
However, at the end of Section \ref{sec:linresp}, we will
present analytical results for the maximum relative
efficiencies achieved when the system operates {\it close to
equilibrium}.

\section{Linear Response}
\label{sec:linresp}

When $f=0$ and $T_A=T_B\equiv\beta^{-1}$, our system is 
in equilibrium, and there
results no average particle drift or heat flow.
In Section \ref{sec:eng_ref} we briefly considered the
behavior of our system {\it near} equilibrium
(see e.g.\ Eq.\ref{eq:ref_lin}).
We now consider this case in more detail.
We will present a general analysis essentially the
same as that of J\" ulicher, Ajdari, and Prost\cite{j},
and then show that the exact results obtained
for our model confirm the predictions of this analysis.

For a sufficiently small load $f$ and inverse temperature
difference $\gamma$, and a fixed value of $\beta$
(characterizing the inverse temperature of the equilibrium
state around which we expand),
we expect to be in the {\it linear
response} regime: the particle drift and heat flows
depend linearly on $f$ and $\gamma$.
Let us introduce the quantity
\begin{equation}
\Phi = {1\over 2} \Bigl( \dot Q_A + \dot Q_B \Bigr)
\end{equation}
-- roughly, a ``heat flux'' from $A$ to $B$ --
and let us write, to leading order
in $f$, $\gamma$:
\begin{equation}
\left(
\begin{array}{c}
v \\ \Phi
\end{array}
\right) =
\left(
\begin{array}{cc}
\partial v/\partial f   ~~&~~ \partial v/\partial\gamma \\
\partial\Phi/\partial f ~~&~~ \partial\Phi/\partial\gamma
\end{array}
\right)
\left(
\begin{array}{c}
f \\ \gamma
\end{array}
\right),
\end{equation}
with the derivatives of $v$ and $\Phi$
evaluated at equilibrium ($f=\gamma=0$).
As per the arguments given at the end of the previous
section, $\dot Q_A$ and $\dot Q_B$ are equal, to leading
order in $f$ and $\gamma$, near equilibrium.

The rate of entropy production is then:
\begin{eqnarray}
\dot S &=& -{\dot Q_A\over T_A} + {\dot Q_B\over T_B} \nonumber\\
&=& -\beta\dot W + \gamma\Phi \nonumber\\
&=&
(f~~\gamma)
\left(
\begin{array}{cc}
M_{11} & M_{12} \\ M_{21} & M_{22}
\end{array}
\right)
\left(
\begin{array}{c}
f \\ \gamma
\end{array}
\right),
\end{eqnarray}
where
$M_{11}=-\beta~\partial v/\partial f$,
$M_{12}=-\beta~\partial v/\partial\gamma$,
$M_{21}=\partial\Phi/\partial f$, and 
$M_{22}=\partial\Phi/\partial\gamma$,
evaluated at equilibrium.
The Second Law implies that
\begin{equation}
\label{eq:det}
\det{\bf M}\ge 0,
\end{equation}
whereas Onsager's reciprocity relation\cite{onsager}
predicts that $M_{12}=M_{21}$, or
\begin{equation}
\label{eq:ons}
-\beta
{\partial v\over\partial\gamma}\Biggr\vert_{f=\gamma=0} =
{\partial\Phi\over\partial f}\Biggr\vert_{f=\gamma=0}.
\end{equation}
Also, the diagonal elements of ${\bf M}$ must be positive:
the particle must slide {\it down} the potential
slope when the temperatures are equal
($M_{11}>0$), and there must be a flow of
heat from the hotter to the colder reservoir when
the slope is zero ($M_{22}>0$).
J\" ulicher {\it et al.}\cite{j} have obtained
identical results for a molecular motor driven
by a difference in chemical potential rather than
temperature.

Using the exact results obtained in Section \ref{sec:load},
we differentiate $v$ and $\Phi$ with respect to $f$ and
$\gamma$ to
evaluate the elements of the matrix ${\bf M}$ for our
model\footnote{
Recall that the expressions for $v({\bf x})$,
$\dot Q_A({\bf x})$, and $\dot Q_B({\bf x})$ differ
according to the sign of $f$.
We have verified that, regardless of whether we use
the results valid for range (b) ($f<0$), or those for
range (c) ($f>0$), we obtain the same results for
the elements of ${\bf M}$.
}:
\begin{equation}
\label{eq:m}
{\bf M} =
\left(
\begin{array}{cc}
3\beta^2d^2C(3+4\zeta) ~&~
\alpha\beta dC(1-\zeta) \\
\alpha\beta dC(1-\zeta) ~&~
\alpha^2C[4+\zeta(29+\zeta)]/(4+3\zeta)
\end{array}
\right),
\end{equation}
where
\begin{equation}
\zeta = e^{-\alpha\beta} \qquad
(0<\zeta<1) \qquad,\qquad
C = {3\zeta\Gamma\over(16+5\zeta)[1+\zeta(4+\zeta)]} > 0.
\end{equation}
The variable $\zeta$ is akin to $\mu$ and $\nu$: it is
the ``rescaled temperature'' of the equilibrium state
with respect to which the linear response behavior is
defined.
Then
\begin{equation}
\det {\bf M} =
{9(\alpha\beta\zeta d\Gamma)^2
(2+19\zeta+21\zeta^2) \over
(4+3\zeta)(16+5\zeta)[1+\zeta(4+\zeta)]^2},
\end{equation}
and we see by inspection that $M_{11},M_{22}>0$;
that $M_{12}=M_{21}$, as mandated
by Onsager reciprocity (Eq.\ref{eq:ons}); and that
$\det {\bf M} >0$, in agreement with the Second Law
(Eq.\ref{eq:det}).

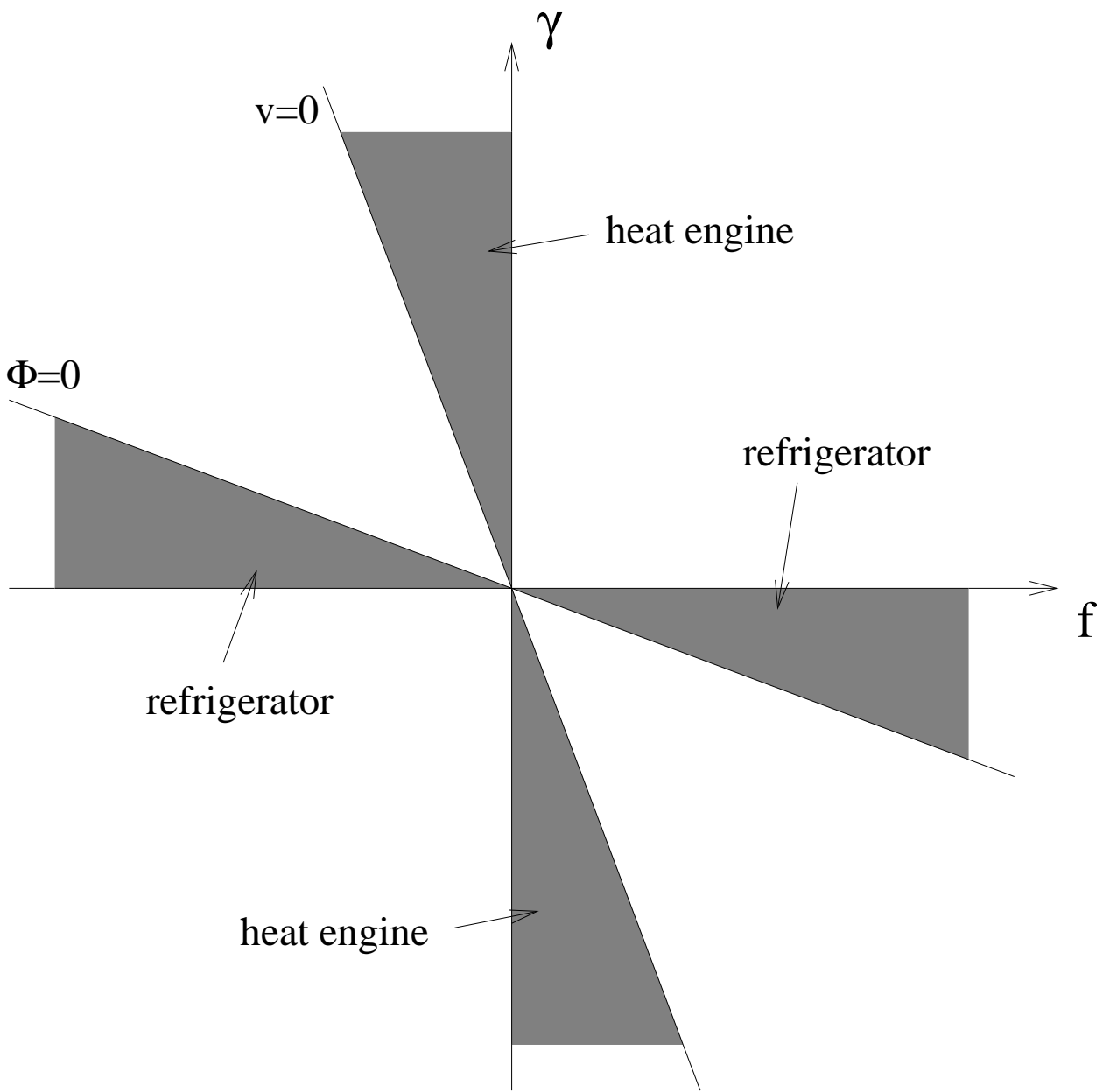
\begin{figure}
\caption{General predictions based on linear
response.
The shaded regions adjacent to the vertical axis
indicate the values of $(f,\gamma)$ for which
the system behaves as a heat engine;
those adjacent to the horizontal axis indicate
that the system is a refrigerator.
The diagonal lines bounding these regions are
the contours $v(f,\gamma)=0$ and $\Phi(f,\gamma)=0$.
This figure essentially combines Figs.\ref{fig:engine}
and \ref{fig:refrig}, for near-equilibrium values
of $f, \gamma\approx 0$.}
\label{fig:linresp}
\end{figure}

It is interesting to consider the operation of our
system as a heat engine and refrigerator, in the
linear response regime.
In this regime, the conditions for these two behaviors
are:
$vf>0$ for a heat engine (as before), and
$\gamma\Phi<0$ for a refrigerator
(since $\Phi=\dot Q_A=\dot Q_B$, to leading order
in $f$, $\gamma$).
In Fig.\ref{fig:linresp}, the shaded regions indicate
values of $(f,\gamma)$ for which the system acts as
a heat engine or refrigerator (compare with Fig.[2]
of Ref.\cite{j}.)
The two diagonal lines which form boundaries of
these regions are given by $v=0$ and $\Phi=0$.
The slopes of these lines are:
\begin{equation}
\lambda_{v=0} = -{M_{11}\over M_{12}} \qquad,\qquad
\lambda_{\Phi=0} = -{M_{21}\over M_{22}}.
\end{equation}
The Second Law, by requiring that $\det {\bf M}\ge 0$,
guarantees that these shaded regions do not overlap:
$\vert\lambda_{v=0}\vert \ge \vert\lambda_{\Phi=0}\vert$.

The {\it efficiency} of our system, when operating
as a heat engine, is given by:
\begin{equation}
\eta_{\rm eng} = {\dot W\over\left\vert\Phi\right\vert}
\end{equation}
(again using $\Phi=\dot Q_A=\dot Q_B$ to leading order).
The {\it Carnot efficiency} defined for the temperatures
$T_A$ and $T_B$ is:
\begin{equation}
\eta_{\rm eng}^C =
{\left\vert T_A-T_B\right\vert\over T_>}
= {\left\vert\gamma\right\vert\over\beta} + O(\gamma^2).
\end{equation}
Then we can get an explicit expression for the
relative efficiency ($y=\eta/\eta^C$) of our
system, in the linear response regime:
\begin{equation}
\label{eq:releff_eng}
y_{\rm eng} \equiv
{\eta_{\rm eng}\over \eta_{\rm eng}^C} =
-{1\over\lambda}
{M_{11}+M_{12}\lambda \over M_{21}+M_{22}\lambda},
\end{equation}
where $\lambda=\gamma/f$.
The relative efficiency is the same for all points along
any straight line through the origin, and Eq.\ref{eq:releff_eng}
gives that relative efficiency as a function of the
slope $\lambda$ of the line.
A similar analysis holds for the case of refrigeration:
\begin{equation}
\label{eq:releff_ref}
y_{\rm ref} \equiv
{\eta_{\rm ref}\over \eta_{\rm ref}^C} =
-\lambda
{M_{21}+M_{22}\lambda \over M_{11}+M_{12}\lambda}.
\end{equation}
Note that $y_{\rm ref}$ happens to be the inverse
of $y_{\rm eng}$, although the two expressions are
valid for different ranges of $\lambda$ values,
corresponding to the shaded regions in
Fig.\ref{fig:linresp}.

The results of the previous two paragraphs were
derived with the implicit assumption that
$M_{12},M_{21}>0$.
This happens to be true for our model, but in
general these elements can be either positive
or negative (or zero), so long as they are equal.
If $M_{12}$ and $M_{21}$ were negative, then
the shaded regions would occur in the the
first and third quadrants of the $(f,\gamma)$
plane, and the negative signs would not appear in
Eqs.\ref{eq:releff_eng} and \ref{eq:releff_ref}.

The above results imply that, near equilibrium,
a microscopic device operating between two
reservoirs either can act {\it both} as a heat engine
and as a refrigerator (if $M_{12}=M_{21}\ne 0$),
or will exhibit neither behavior (if $M_{12}=M_{21}=0$).
For instance, in a microscopic ratchet-and-pawl
device, if the sawteeth on the wheel have a symmetric
shape, then the system cannot behave as a heat engine:
$M_{12}=0$, as is obvious by symmetry.
What is not so immediately obvious, but follows from
the condition $M_{12}=M_{21}$, is that it is
equally impossible for a system with symmetric teeth
to operate as a refrigerator, in the linear response
regime.

Finally, for a given inverse temperature $\beta$ of the
equilibrium state, we can solve for the {\it maximal}
relative efficiency achievable in the linear
response regime, by maximizing
$y_{\rm eng}$ and $y_{\rm ref}$ with respect to
$\lambda$.
It turns out that the maximal efficiencies are
equal in the two cases, and depend only on a single
parameter $r$ characterizing the behavior of the
system near equilibrium:
\begin{equation}
y^{\rm MAX} \equiv
y_{\rm eng}^{\rm MAX} = 
y_{\rm ref}^{\rm MAX} =
{r\over (1 + \sqrt{1-r})^2}
\qquad,\qquad r(\beta) = {M_{12}M_{21}\over M_{11}M_{22}}
\qquad (0\le r\le 1).
\end{equation}
The value of $y^{\rm MAX}$ increases monotonically as $r$ 
goes from 0 to 1.

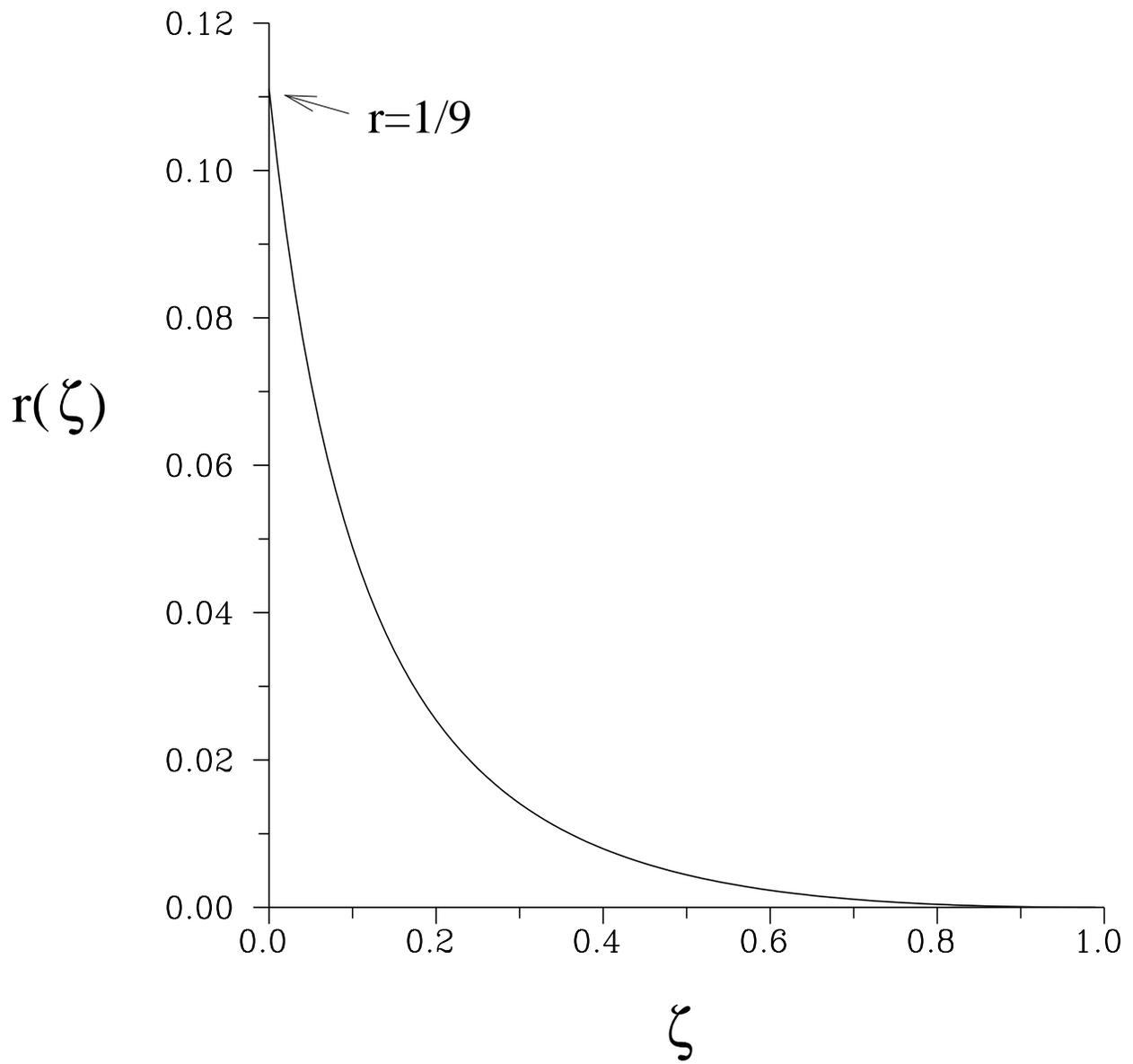
\begin{figure}
\caption{The quantity $r(\zeta)$ is plotted for the
entire range of values of the (rescaled) 
equilibrium state temperature $(0<\zeta<1)$.}
\label{fig:r}
\end{figure}

From Eq.\ref{eq:m}, we find that our model gives:
\begin{equation}
r =
{(1-\zeta)^2(4+3\zeta) \over
3(3+4\zeta)[4+\zeta(29+9\zeta)]},
\end{equation}
which depends only on the rescaled temperature
$\zeta=\exp(-\alpha\beta)$ of the equilibrium state.
In Fig.\ref{fig:r} we plot $r(\zeta)$.
We see that $r$ approaches a limiting value of 1/9
as the equilibrium temperature $T=\beta^{-1}$ goes
to zero (i.e.\ $\zeta\rightarrow 0$), and decreases
to zero as $T\rightarrow\infty$ (i.e.\ $\zeta\rightarrow 1$).
For the limiting value $r=1/9$, our model gives
a maximal relative efficiency
\begin{equation}
\label{eq:maxreleff}
y^{\rm MAX}(T\rightarrow 0) =
{1\over 17 + 12\sqrt 2} \approx 0.0294.
\end{equation}
This is the best relative efficiency which our system
can achieve near equilibrium.

We have carried out a cursory numerical search --
$10^8$ points sampled randomly in $(f,\mu,\nu)$-space --
and have found, {\it away from equilibrium}, 
relative efficiencies as high as
$y\approx 0.0432$ (heat engine) and
$y\approx 0.0647$ (refrigerator).
These are greater than the near-equilibrium value quoted in
Eq.\ref{eq:maxreleff}, but still far short of unity.
This suggests that
the efficiency of our model is always considerably
lower than the corresponding Carnot efficiency.
Such a conclusion is
in agreement with Parrondo and Espanol\cite{n},
and Sekimoto\cite{o}, who have argued that
Feynman's analysis on the point of efficiency --
in which he concluded that the ratchet and pawl would
operate at Carnot efficiency -- was in error.
We note also that Hondou and Takagi\cite{p},
as well as Magnasco and Stolovitzky\cite{q},
have shown that, within a Langevin
model for two degrees of freedom (e.g.\ the ratchet
and pawl) coupled to two different heat reservoirs,
Carnot efficiency cannot be achieved.

\section{Summary}

Our aim in this paper has been to introduce a discrete model
of Feynman's ratchet and pawl, and to solve exactly for the
behavior of that model as a function of external load ($f$)
and reservoir temperatures ($T_A$, $T_B$).
The central result, Eqs.\ref{eq:ss}, gives the average directed
motion ($v$) and heat flows ($\dot Q_A$, $\dot Q_B$), in
the steady state.
We have shown that our model can act both as a heat engine
and as a refrigerator, and we have investigated its
behavior in the near-equilibrium, linear response regime.

\section{Acknowledgments}

We would like to thank Dr.\ Martin Bier for stimulating
correspondence regarding thermal ratchets, for stressing
that our model is a discrete version of Feynman's ratchet 
and pawl, and for his 
clear lecture at the 1998 Marian Smoluchowski Symposium 
on Statistical Physics (Zakopane, Poland), which
introduced us to this interesting topic.
We would also like to thank Drs.\ Shankar Subramaniam
and Wojciech Zurek for illuminating discussions regarding
thermodynamic efficiencies.
This work was carried out during reciprocal visits to 
Warsaw and Los Alamos, made possible by financial
support from the
Polish-American Maria Sk\l odowska-Curie Joint
Fund II, under project PAA/NSF-96-253,
as well as by the hospitality of the Institute
for Nuclear Studies (Poland) and Los Alamos National
Laboratory (USA).

\section*{Appendix}
\label{sec:appendix}

Here we present explicit expressions for the 
polynomials $P_n^j$, $X^j$, and $Y^j$ appearing
in Section \ref{sec:load}.
The $P_n^j$'s are obtained by Jordan elimination, 
performed on the matrices ${\cal R}^j$, $j=b,c$; and
$X_j$ and $Y_j$ then follow from 
Eqs.\ref{eq:vj},\ref{eq:jpm},\ref{eq:qadot},\ref{eq:qbdot},
\ref{eq:pnj}, and \ref{eq:ss}.

\begin{eqnarray}
%
%
%
P^b_1 &=&   \nu^3 \sigma (1 + \sigma + 3 \sigma^2) + 
       \nu \sigma^2 (4 + 3 \mu + (4 + \mu) (1 + 2 \mu) \sigma 
+ (4 + 3 \mu (3 + 2 \mu)) \sigma^2)
\nonumber\\ & &         
       + 4 \mu \sigma (3 + \sigma (5 + 3 \mu + 5 \sigma 
+ 4 \mu \sigma)) + 
       \nu^2 (4 + \sigma (4 + 2 \mu + \sigma (6 + \sigma 
+ \sigma^2 + 3 \sigma^3)))
\nonumber\\
P^b_2 &=&    \nu \sigma^2 (4 + 3 \mu + (12 + 7 \mu) \sigma + 
(12 + \mu (15 + 2 \mu)) \sigma^2) + 
       \nu^2 (4 + 2 (6 + \mu) \sigma + 6 (1 + \mu) \sigma^2 
\nonumber\\ & &         
       + \sigma^3 + 3 \sigma^4 + \sigma^5) + 
       \nu^3 \sigma (1 + \sigma (3 + \sigma)) 
       + 4 \mu \sigma (3 + \sigma (3 + \mu + 3 (1 + \mu) \sigma))
\nonumber\\
P^b_3 &=&    4 \mu \sigma (3 + \sigma (5 + 3 \sigma)) 
+ \nu^3 \sigma (1 + \sigma (5 + 9 \sigma)) + 
       \nu \sigma^2 (4 (1 + \sigma + 3 \sigma^2) 
\nonumber\\ & &         
+ \mu (3 + \sigma (9 + 7 \sigma))) + 
       \nu^2 (4 + \sigma (12 + \sigma (18 + \sigma 
+ 5 \sigma^2 + 9 \sigma^3)))
\nonumber\\
P^b_4 &=&    \sigma (16 + 4 (6 + \nu) \sigma + 6 (4 + \nu) \sigma^2 
+ 10 \nu \sigma^3 + 
          \mu (4 + \sigma (20 + \nu + 24 \sigma 
+ 5 \nu \sigma + 9 \nu \sigma^2)))
\nonumber\\
P^b_5 &=&    \nu^2 (4 + \mu + 5 (2 + \mu) \sigma 
+ 3 (2 + 3 \mu) \sigma^2) + 
   2 \mu \sigma (4 + \mu + (4 + 3 \mu) \sigma 
+ (4 + 5 \mu) \sigma^2) 
\nonumber\\ & &         
 +   2 \nu \sigma^2 (4 + \mu + (8 + 5 \mu) \sigma 
+ (8 + 9 \mu) \sigma^2)
\nonumber\\
P^b_6 &=&    2 \mu \nu \sigma^2 (2 + \mu 
+ (2 + 3 \mu) \sigma + (4 + 3 \mu) \sigma^2) 
+ 2 \mu^2 \sigma (3 + \sigma (5 + 3 \sigma)) 
\nonumber\\ & &
+ \nu^3 \sigma (2 + 4 \sigma (1 + \sigma) 
+ \mu (1 + \sigma (5 + 9 \sigma))) + 
     \nu^2 (4 + 6 \sigma + 2 \sigma^2 (3 + \sigma (1 
+ 2 \sigma (1 + \sigma)))
\nonumber\\ & &         
      + \mu (3 + \sigma (11 
      + \sigma (15 + \sigma 
+ 5 \sigma^2 + 9 \sigma^3))))
\\ & & \nonumber\\
%
%
%
X^b &=&    \nu^3 \sigma (-1 - 2 \sigma + 3 \sigma^3) + 
       \nu \sigma^2 (-4 - 3 \mu - 2 \mu \sigma 
       + 4 \mu \sigma^2 + 3 (4 + \mu (5 + 2 \mu)) \sigma^3) 
\nonumber\\ & &         
       +  \nu^2 (-4 - 3 (4 + \mu) \sigma - 3 (4 + 3 \mu) \sigma^2 
       - (1 + 9 \mu) \sigma^3 + (2 + \mu) \sigma^4 + 
          (6 + 5 \mu) \sigma^5  
\nonumber\\ & &         
	+ (13 + 9 \mu) \sigma^6)
          +  2 \mu \sigma (-6 + \sigma (-4 + 6 \sigma^2 
       + 3 \mu (-1 + (-1 + \sigma) \sigma)))
\nonumber\\
%
%
%
Y^b &=&    - (\nu^3 \sigma (3 + \sigma (5 + 7 \sigma)) + \nu \sigma^2 
       (-2 \mu^2 (1 + \sigma) (1 + 2 \sigma) 
       + \mu (-1 + (3 - 5 \sigma) \sigma) 
\nonumber\\ & &         
       + 4 (1 + \sigma + \sigma^2)) - 
       2 \mu \sigma (2 (1 + \sigma + \sigma^2) + \mu (5 
       + \sigma (9 + 7 \sigma))) 
\nonumber\\ & &         
       +  \nu^2 (8 + \mu (-1 + \sigma - 3 \sigma^2) 
      + \sigma (10 + \sigma (12 + \sigma (3 
      + \sigma (5 + 7 \sigma))))))
\\ & & \nonumber\\ & & \nonumber\\
P^c_1 &=& \nu^3 \sigma (1 + \sigma (3 + \sigma)) + 
4 \mu \sigma (5 + \mu (5 + 2 \sigma) + \sigma (5 + 3 \sigma)) + 
\nu \sigma^2 (2 \mu^2 (3 + \sigma) \nonumber\\
& &+ 4 (1 + \sigma + \sigma^2) +3 \mu (3 + \sigma (3 + \sigma))) + 
\nu^2 (4 + 2 \mu + \sigma (6 + \sigma 
(4 + \sigma + 3 \sigma^2 + \sigma^3)))
\nonumber\\
P^c_2 &=& \nu^3 \sigma (3 + \sigma + \sigma^2) + 
4 \mu \sigma (\mu + 3 \mu \sigma + 3 (1 + \sigma + \sigma^2)) + 
\nu \sigma^2 (2 \mu^2 \sigma + 4 (3 + \sigma (3 + \sigma))
\nonumber\\ & &         
 + \mu (15 + \sigma (7 + 3 \sigma))) + 
\nu^2 (12 + 2 \mu (3 + \sigma) + 
\sigma (6 + \sigma (4 + \sigma (3 + \sigma + \sigma^2))))
\nonumber\\
P^c_3 &=&  4 \mu \sigma (5 + 3 \sigma (1 + \sigma)) + 
\nu^3 \sigma (9 + \sigma (5 + \sigma)) + 
\nu \sigma^2 (4 (1 + \sigma (3 + \sigma)) 
\nonumber\\ & &         
+ \mu (9 
+ \sigma (7 + 3 \sigma))) + 
\nu^2 (16 + \sigma (14 + \sigma (4 + 
\sigma (9 + \sigma (5 + \sigma)))))
\nonumber\\
P^c_4 &=&  \sigma (24 + 2 \sigma (3 + 2 \sigma) 
(4 + \nu + \nu \sigma) +  \mu (28 + \sigma (4 (4 + \sigma) 
+ \nu (9 + \sigma (5 + \sigma)))))
\nonumber\\
P^c_5 &=&  2 \mu \sigma (4 (1 + \sigma + \sigma^2) + 
\mu (3 + \sigma (5 + \sigma))) +  \nu^2 (10 + 6 \sigma 
+ 4 \sigma^2 + \mu (9 + \sigma (5 + \sigma)))
\nonumber\\ & &         
+2 \nu \sigma^2 (4 (2 + \sigma (2 + \sigma)) 
+ \mu (9 + \sigma (5 + \sigma)))
\nonumber\\
P^c_6 &=&       2 \mu^2 \sigma (5 + 3 \sigma (1 + \sigma)) + 
2 \mu \nu \sigma^2 (2 (1 + \sigma)^2 + \mu (3 + \sigma (3 + \sigma))) 
+ \nu^3 \sigma (2 (2 
+ \sigma (2 + \sigma)) 
\nonumber\\ & &         
+ \mu (9 + \sigma (5 + \sigma))) + 
\nu^2 (2 (3 + \sigma (3 + \sigma (2 + \sigma (2 + \sigma (2 + \sigma))))) 
\nonumber\\ & &         
+ \mu (15 + \sigma (11 + \sigma (3 + \sigma (9 + \sigma (5 + \sigma))))))
\\ & & \nonumber\\
%
%
%
X^c &=&   \sigma (\nu \sigma^2 (-4 - 5 \mu + 4 \mu (2 + \mu) 
\sigma + 2 (2 + \mu)^2 \sigma^2 + 
(4 + 3 \mu) \sigma^3) + \nu^3 (-3 \sigma + 2 \sigma^3  + \sigma^4) 
\nonumber\\ & &         
+  2 \mu \sigma (-10 + 6 \sigma^3 
+ \mu (-5 + \sigma (-1 + 3 \sigma))) 
\nonumber\\ & &         
+ \nu^2 (-22 + \mu (-15 + \sigma (-5 + \sigma (-1 + \sigma (9 
  + \sigma (5 + \sigma))))) 
\nonumber\\ & &
  + \sigma (-8 + \sigma (-2 + \sigma (7 + \sigma (10 
  + \sigma (6 + \sigma)))))))
\nonumber\\
%
%
%
Y^c &=&   -(\nu^3 \sigma (5 + \sigma (7 + 3 \sigma)) + 
         \nu \sigma^2 (4 (1 + \sigma + \sigma^2) 
         - 2 \mu^2 (3 + \sigma (2 + \sigma)) 
\nonumber\\ & &         
- \mu (-3  
+ \sigma (5 + \sigma))) -         
2 \mu \sigma (2 (1 + \sigma + \sigma^2) 
         + \mu (9 + \sigma (7 + 5 \sigma))) 
\nonumber\\ & &         
+ \nu^2 (10 - \mu (-1 + \sigma (3 + \sigma)) 
         + \sigma (12 + \sigma (8 + \sigma (5 + \sigma (7 + 3 \sigma)))))) 
\end{eqnarray}

%
%

\end{document}